# Treatment Geometry and Causal Identification with Earth Observation Data

Jeffrey D. Michler, Anna Josephson, Elinor Benami, Patrick Behrer, Michael J. Cecil, Sydney Gourlay, Robert Heilmayr, Ella Kirchner, Gina Maskell, Kunwar Singh

**Abstract**

A central task in conducting impact evaluations is determining who or what was exposed to a treatment, when, and to what degree. These questions can be especially complex in geospatial settings, where many reasonable definitions of exposure may exist. This chapter introduces treatment geometry as a core concept in geospatial impact evaluation (GIE): the spatial and temporal footprint of a treatment as represented in data. How this footprint is defined shapes identification strategies and the credibility of causal inference. Drawing on cases spanning the air pollution, wildfire, forest policy, infrastructure, pest, and food security literature, the chapter provides practical guidance on navigating key tradeoffs (including spatial resolution, temporal alignment, spillovers, and boundary uncertainty) that arise when translating real-world interventions into analyzable data. Rather than prescribing a single best approach, the chapter equips researchers with a framework for diagnosing which geometry decisions may matter most in their context, closing with synthesis questions to help readers navigate these decisions.

**Learning objectives:**

1. Define and justify the spatial and temporal footprint of treatment in a geospatial impact evaluation.
2. Assess trade-offs in treatment geometry choices, including scale, timing, intensity, spillovers, and uncertainty.
3. Integrate EO-derived data with socioeconomic or administrative datasets to construct reproducible treatment and/or outcome measures.

**Table of Contents**

# 1. Introduction

Geospatial impact evaluation (GIE) often involves merging data from remotely-sensed Earth observation (EO) products with socioeconomic data to answer causal questions on the impacts and effectiveness of a variety of human interventions or natural environmental events. The resulting research design may regress socioeconomic data on EO data (e.g., how drought affects irrigation investments); EO on socioeconomic (e.g., how Indigenous tenure reform affects land use); and EO on EO (e.g., how temperature affects crop yields) (Donaldson and Storeygard, 2016; Hsiang, 2016; Jain, 2020). In some cases, the spatial and temporal alignment may be straightforward, particularly when aggregated over larger spatial areas, such as global analyses that aggregate at the national level or when pixels can be statistically summarized into country boundaries. A bigger challenge emerges when the socioeconomic sampling design and the resolution of EO products differ in such a way that one cannot easily align the spatial and/or temporal features of the various types of data. Regardless of the implementation, a core element of the research design is defining who or what was "treated," when treatment began and ended, and what constitutes a credible comparison ("control") unit.

To systematically operationalize this element of research design, we introduce the concept of a treatment geometry. The term originates in radiation oncology, where it refers to the spatial orientation of a tumor within a body and the resulting alignment of patient, organs at risk, and radiation beams that balances the tradeoff between covering the tumor and sparing normal tissue so as to deliver the most effective treatment dose (Bortfeld, 2006). In the context of GIEs, we use treatment geometry to refer to the spatial and temporal footprint through which exposure to treatment is represented. This encompasses the polygons, buffers, directionality, and time windows used to encode exposure as well as the design choices a researcher makes to align EO-derived measures with socioeconomic data given their study context.

Treatment geometry is inherently an analytical choice to be justified by the underlying causal data generating processes of a given context (Goodchild, 1992; Kwan, 2012). Different interventions have different spatiotemporal patterns of influence: effects may occur within defined boundaries (e.g., protected areas), follow specific directions (e.g., downstream hydrology), move along geographic features (e.g., transportation corridors), or diffuse across space as a function of time (e.g., storm paths). Treatments may be discrete or fuzzy, short-term or cumulative, and their intensity may vary both spatially and temporally. Explicitly defining treatment geometry forces transparency about assumptions that are often left implicit, including the assumed shape of influence, the boundary at which exposure becomes "zero," and how exposure maps onto observed units. For example, policy interventions tied to administrative boundaries may spill over into neighboring jurisdictions if markets, mobility, or labor respond to prices.

Multiple reasonable definitions may exist for the same empirical question, each with their own tradeoffs. A deep understanding of the spatial processes that define a treatment geometry improves measurement of dose-response relationships, reduces misclassification bias, and clarifies the scale and scope of estimated effects. It can also be used to justify research designs (e.g., spatial discontinuities) or test core assumptions (e.g., the absence of spillovers), ultimately enhancing the credibility, transparency, and interpretability of GIEs across diverse interventions and environmental contexts.

The remainder of this chapter focuses on the design choices facing a researcher and how the treatment geometry concept helps surface the trade-offs embedded in each choice. Section 2 defines a treatment geometry as a spatial and temporal footprint through which exposure to a policy, intervention, hazard, or environmental circumstance is represented. We show how researchers can operationalize the concept by decomposing treatment geometry into its source, treated unit, pathway, and temporal window. Section 3 discusses how to develop a protocol or assignment rule to translate "treatment" into variables necessary for analysis. This includes both an empirical measure of treatment, such as a dummy for if a weather shock occurred or a continuous measure of pm2.5 levels, and assigning that measure to a unit of analysis, such as a household, grid cell, or administrative unit. Section 4 elucidates how research design choices in defining treatment geometry ease or threaten causal identification and how researchers can build confidence in their causal nature of their estimates. Section 5 presents a set of case studies to illustrate how these issues arise in applied research, covering topics from spillovers and transport pathways to data sparsity, treatment timing, and ground data registration. Section 6 concludes with a treatment geometry protocol that summarizes good and better practices for defining, documenting, visualizing, testing, and reproducing treatment assignments in GIE.

# 2. Operationalizing Treatment Geometry

Operationalizing treatment geometry requires more than just a mechanical overlay of one dataset on another (Gotway and Young, 2002). It is an empirical research design choice that links a substantive theory of exposure to a measurable variable. It requires making empirical choices about what the treatment is, where its effects are expected to occur, when those effects are relevant, how exposure should be assigned to observed units, and how uncertainty in those choices should be handled. Each choice creates a fork in the path and the decision to follow one trajectory instead of a plausible alternative path can affect identification of causal effects.

The central task is to construct a defensible mapping from the underlying unobservable process to the observed data (Goodchild, 1992; Gotway and Young, 2002). Each design choice may be analytically separable but are often intertwined in practice. The choice of a spatial buffer may also imply a temporal window; the choice of a seasonal exposure definition may change which spatial units are treated; and the choice to aggregate pixels to districts may change whether treatment is interpreted as household exposure, local average exposure, or jurisdiction-level exposure. For this reason, treatment geometry should be defined early in the research design and revisited throughout the analysis. Throughout we tend to frame examples as cases where treatment comes from EO data and outcomes from socioeconomic data, but the same lessons apply when the reverse is true or when both come from EO data.

## 2.1. Decomposing Treatment Geometry

A useful way to operationalize the concept of treatment geometry is to decompose it into a small number of analytical design components. At minimum, researchers should specify four elements: (1) **the source** of treatment exposure; (2) **the unit** that is treated; (3) **the pathway** through which treatment is assigned or travels; and (4) **the temporal** window relevant to treatment. Together these elements define the conceptual geometry of treatment which is then mapped or translated into a treatment variable through an explicit exposure assignment protocol.

<u>The source of treatment</u> is the object, event, policy, or condition from which treatment originates. It may be a fixed point that is the literal source of treatment, such as a factory, mine, power plant, health facility, school, or conflict event. In other cases, the source may be a bounded footprint or polygon that defines where a policy, intervention, hazard, or environmental condition applies. These may be protected areas, titled land parcels, administrative jurisdictions, flood extents, fire perimeters, or crop-stress zones. Or, the source of treatment may be a line or network, such as a road, river, canal, transmission line, or transport corridor. It may also be a continuous surface, such as rainfall, temperature, smoke, vegetation health, drought severity, or nighttime lights (Goodchild, 1992). The source determines what kind of spatial object the researcher begins with, but it does not by itself determine who is treated.

The unit of treatment is the unit record that receives the treatment. These may be households, firms, plots, parcels, schools, health facilities, villages, enumeration areas (EAs), administrative units, pixels, or grid cells. Treatment geometry is inherently about the relationship between the source of treatment and the "receiving" units. For example, a protected-area polygon may define the boundary of a conservation intervention, but the treated units are forest pixels, farms near the boundary, communities adjacent to the park, or administrative units overlapping the protected area. Similarly, a flood polygon may describe the spatial extent of an event, but treatment may be assigned to households, roads, parcels, or crop fields that intersect that footprint. In other cases, the protected-area polygon that defines treatment status may itself be the unit of treatment if the outcome is measured at the protected-area level. The same spatial feature can thus play different roles depending on the estimand and level of analysis. Whether the source of treatment and the unit of treatment are distinct objects or the same object depends on the research design.

The pathway of treatment describes how treatment is expected to move through space. Some treatments are spatially bounded by design, as with policies applied within administrative units or land parcels. Others decay with distance, as may occur with some localized environmental hazards or infrastructure effects. Others move directionally, as with air pollution, wildfire smoke, water pollution, flood propagation, or pests. Still others move through social, market, or ownership networks rather than geographic proximity alone. In these cases, distance may be a poor proxy for exposure unless the pathway is explicitly incorporated into the geometry.

The temporal window of treatment defines when treatment is relevant. Some exposures are acute, such as a tropical cyclone, fire, flood, or conflict event. Others extend or even accumulate over time, such as soil degradation, the duration of a drought, wildfire smoke exposure, or the length of a war. Some treatments begin at a clearly defined date, while others involve anticipation, delayed implementation, incomplete enforcement, or gradual adoption. The temporal window should correspond to the mechanism being tested rather than simply to the frequency of available data. Annual measures may be appropriate for some questions, but they may be too coarse for shocks whose impacts vary based on their timing within the year (e.g. if they depend on crop stage, season, or short-run behavioral responses).

Together, the source, unit, pathway, and temporal window define the conceptual geometry of treatment. They clarify why alternative geometries can be reasonable for the same broad research question: each definition captures a different causal pathway and supports a different estimand. The goal is not to find a universally correct geometry but to choose and justify the geometry that best matches the mechanism, outcome, and data.

## 2.2. Representing Treatment

The first set of design choices for representing treatment focuses on the exposure, scale, timing, and intensity of the treatment. These choices determine which units are classified as treated, which units are classified as untreated, and what variation is used for identification. They also determine how much spatial and temporal heterogeneity is preserved or lost as EO and socioeconomic data are merged.

### 2.2.1. Spatial and Temporal Footprints

The spatial footprint of treatment is the area over which treatment is expected to affect outcomes (Kwan, 2012). This footprint may coincide with a formal boundary but often does not. A protected area has a legal boundary yet its effects may extend outside that boundary through leakages like conservation spillovers, tourism, or market responses. A road has a physical footprint, but its economic effects may extend to nearby firms, farms, labor markets, and settlements. A pollutant source has a point location, but exposure depends on how pollutants disperse through the atmosphere or water system. A locust outbreak, cyclone, or flood may be observed at a specific time and place, but its socioeconomic consequences depend on what assets, crops, households, or markets lie within the affected area.

The temporal footprint is equally important. It defines when exposure begins, how long it lasts, and whether effects are expected to be immediate, lagged, cumulative, or reversible (Schlenker and Roberts, 2009; Hsiang, 2016). For some treatments, onset is clear: a storm makes landfall, a fire begins, a road opens, or a policy is implemented. For others, onset is gradual or ambiguous: enforcement ramps up, households learn about a program, drought conditions accumulate, or vegetation stress emerges over time. Some treatments are irreversible, such as clearing a primary forest, while others can be undone, such as the receding of floodwaters. For other treatments, the relevant end date may be difficult to define. Fires may be quickly extinguished while damage persists; smoke exposure may last only days while health impacts persist for weeks; a land tenure reform may be implemented at one date while land-use responses unfold over years.

Defining the footprint therefore requires more than locating the treatment in space and time. It requires specifying the period and area over which the treatment could plausibly affect the outcome of interest. The relevant footprint for crop yields may be different from the footprint for household income, food security, migration, or asset sales. The same shock may affect these outcomes through different pathways and over different time horizons.

### 2.2.2. Resolution and Aggregation

Fine spatial resolution can preserve meaningful variation in treatment, particularly when exposure is heterogeneous within administrative units or when outcomes are tied to specific parcels, plots, or assets. However, finer resolution data is not always an improvement. High-resolution exposure data can magnify georegistration errors, boundary uncertainty, classification noise, or mismatch between the observed location and the unit actually affected. If household coordinates are displaced, if plot boundaries are approximate, or if the EO product is noisier at fine resolution due to modeling, assigning treatment at the finest available scale may create false precision.

Coarser aggregation can reduce noise and align more naturally with policy or outcome units but it can also smooth away the variation needed for inference. District-level averages may be reasonable for policies implemented at the district level but they may mask within-district differences in rainfall, smoke, elevation, market access, or land cover. Aggregating to administrative units can also create sensitivity to the Modifiable Areal Unit Problem: estimated relationships may change when the same underlying data are summarized using different spatial boundaries (Wong, 2004; Avelino et al., 2016).

Researchers should thus choose the scale of treatment assignment by asking where the mechanism operates and where the outcome is measured. However, there is often a mismatch between the resolution of EO data, which are available as pixels or gridded surfaces, and socioeconomic data, which may be observed at plots, households, firms, villages, or administrative zones. Moving from one spatial support to another is rarely if ever neutral (Blackman et al., 2024). If the mechanism operates at the field level, but the outcome is measured at the district level, the analysis must justify how field-level exposure is summarized. Alternatively, if the mechanism operates through markets or administrative implementation, then aggregating to a market area or jurisdiction may be appropriate even when finer EO data are available. Aggregating pixels to polygons, assigning gridded values to points, or interpolating sparse point observations into a surface imposes assumptions about how exposure varies within the relevant unit. The key is to avoid treating resolution as a purely technical matter.

### 2.2.3. Temporal Alignment and Dynamics

Temporal alignment raises issues analogous to spatial resolution and aggregation. Aggregating over a calendar year is not always the most fitting temporal resolution for cases such as treatments that are relevant. A heatwave or pest outbreak may have very different consequences depending on when it occurs relative to crop growth stages (Schlenker and Roberts, 2009; Gong et al., 2024). A shock during establishment may allow for replanting. Rain during grain filling may have a positive effect on output, storage, market access, and income while rain at harvest will likely have the opposite effect. The same event can therefore have very different outcomes depending on timing and how that event is mapped onto the unit of treatment.

Temporal dynamics also matter. Pollution exposure may be acute, repeated, or cumulative. Infrastructure investment may generate anticipatory effects before completion and adjustment effects after completion. Disasters may have immediate effects on assets and longer-run effects on migration, schooling, health, or labor allocation. These dynamics do not always map to EO data in a symmetric or uniform way. When an electrical grid is turned on it is recorded as "lit" in nightlights data as a binary event (Henderson et al., 2012). But once a place is recorded as lit, the signal may persist in the EO data even as the electrical grid decays, producing more brownouts and blackouts.

Researchers should thus distinguish among exposure onset, exposure duration, response timing, and outcome measurement. As with spatial resolution, there is often a mismatch between the time-step of EO data, which are acquired at daily, weekly, 10-day, or annual intervals, and socioeconomic data, which are retrospective or simply snapshots in time. A common temptation is to aggregate EO data to the same calendar period as the socioeconomic data. For some research questions, this may be appropriate but it can be misleading when the relevant mechanism follows a different temporal structure. A defensible treatment geometry should specify whether treatment is measured as a contemporaneous shock, a lagged exposure, a cumulative dose, a moving average, a period above a threshold, or a sequence of stage-specific indicators.

### 2.2.4. Treatment Intensity and Dose-Response Relationships

Treatment intensity raises the question of whether exposure should be measured as occurrence, severity, duration, or accumulation (Hirano and Imbens, 2005). A binary treatment captures the extensive margin: whether a unit was exposed at all within the relevant treatment window. This can be appropriate when crossing an exposure threshold changes behavior. A single nearby conflict event may alter expectations of violence, even if fatalities are not the main mechanism. Continuous intensity measures that capture the intensive margin are more appropriate when severity, duration, or repeated exposure are central, such as rainfall deficits, heat exposure, smoke concentration, bombing intensity, inundation depth, or vegetation stress.

Dose-response relationships raise a related but distinct challenge. Once intensity is measured, researchers must decide how outcomes are expected to vary with exposure. A continuous specification assumes that marginal changes in the exposure scale are meaningful and comparable across units. This may be plausible for some measures, such as cumulative smoke exposure or bombing density, but less plausible for indices whose units are statistical or context-dependent. For example, drought indices can be used continuously, discretized into categories such as moderate, severe, or extreme drought, or converted into counts of dry periods during a growing season. Each choice identifies a relationship between exposure and outcome and implies assumptions about nonlinearity, thresholds, and comparability.

Choosing between intensity measures and dose-response specifications therefore requires aligning the treatment variable with the causal mechanism, measurement, quality, and estimand. A binary measure estimates the effect of crossing an exposure threshold. A categorical measure allows effects to differ across severity classes. A continuous measure estimates how outcomes change with marginal increases in exposure, conditional on the scale being meaningful. Because thresholds can be arbitrary, researchers should justify them and examine sensitivity to alternatives. When theory suggests nonlinear effects but does not identify a precise cutoff, bins, flexible functional forms, or severity categories may be preferable to a single threshold.

#### 2.2.5. Boundary and Directional Uncertainty

Boundary uncertainty can produce treatment misclassification. Fire perimeters, flood extents, crop maps, land-cover classifications, pollution surfaces, and administrative boundaries are all measured with error (Foody, 2002; Olofsson et al., 2014). A household near a flood boundary may be incorrectly classified as untreated if the flood extent is underestimated or incorrectly classified as treated if the observed boundary includes water that did not affect the household's assets. Survey coordinates in public use data sets are intentionally displaced and may be linked to the wrong rainfall pixel, land-cover class, or pollution surface (Perez-Heydrich et al., 2016; Warren et al., 2015; Michler et al., 2022). The bias from such errors need not be simple attenuation bias. Bias induced by boundary uncertainty depends on whether the resulting misclassification is random, correlated with outcomes, or correlated with other determinants of treatment.

Directional uncertainty is a related but distinct issue. Many treatments do not radiate uniformly from a source. Smoke, water pollution, floods, pests, and some ecosystem services move through directional processes shaped by the natural environment, the built environment, or ecological pathways (Miller et al., 2019; Zabrocki et al., 2022). In these cases, simple distance-based buffers can assign equal treatment to units with very different exposure levels. A road may affect locations connected to the network differently from locations that are geographically close but poorly connected. A farm connected to restored habitat by hedgerows or riparian corridors may receive more pollination or pest-control spillovers than a closer farm separated from that habitat by roads, cleared land, or pesticide-intensive fields.

As a result, defining the boundary or direction of treatment requires more than locating the treatment within a sharp border or along a linear vector. It requires specifying how boundaries and directionality are measured, with what degree of uncertainty, and how coordinates in socioeconomic data are mapped to those boundaries and vectors. Uncertainty matters most when treatment status changes sharply at a boundary or when the outcome is measured near the edge of the treated area. Even when boundaries or directional processes are approximated, this approximation should be explicit and sensitivity to reasonable alternatives should be reported.

# 3. Assigning Exposure

Once the relevant exposure, scale, timing, and intensity of the treatment has been defined, the researcher must construct the treatment geometry in the data. This requires establishing a protocol that maps or translates the conceptually defined treatment into an empirical measure, which involves two related but distinct processes. One is assigning exposure to an empirical variable, such as a dummy that captures whether a household lies within a 5-km buffer of a road or the share of a forest burned by fire. The other process is assigning exposure to a unit of analysis, such as mapping displaced household coordinates to a gridded raster surface or calculating the area weighted average of exposed grid cells within an administrative unit. This assignment rule is where many implicit assumptions tend to enter the analysis. Explicitly stating the assignment protocol clarifies what comparison is being made and what effect the resulting estimate can be interpreted as capturing.

## 3.1. Assigning Exposure to an Empirical Variable

Having chosen the spatial representation of treatment, one next needs to map that to a clearly defined empirical variable (Goodchild, 1992; Gotway and Young, 2002). The assignment protocol allows for consistent mapping from one domain (the conceptual definition of treatment) to another domain (the empirical value of treatment).

When treatment is represented as a point, line, or polygon, the first task is to map that vector feature onto the EO grid or other spatial data structure to be used in the analysis (Goodchild, 1992; Hengl, 2006). When treatment is represented as a point feature (e.g., the location of a factory, a clinic) it can be assigned to the cell containing its coordinates in the gridded EO data and exposure assigned the value of that grid cell (e.g., pm2.5 level, NDVI value). More often the point serves as the anchor for a derived exposure measure, such as a distance surface, a buffer, a count of sources within a radius, or a directional plume. When treatment is represented as a line feature (e.g., a canal, an administrative border) it can be rasterized into every grid cell it crosses, with exposure assigned as line presence, distance to the line, segment within the cell, travel time to the nearest segment, or upstream/downstream position. When treatment is represented as a polygon feature (e.g., a conservation area, a flood footprint) it can be mapped to cells whose centroids fall inside the polygon, cells touched by the polygon, or cells weighted by fractional overlap. A household point may be coded as treated if it lies inside a flood boundary. A field polygon may be coded as treated based on the share of area that overlaps with a crop stress layer. A district may be coded as treated based on the inverse of the distance to a rail line. The relevant matching rule should be matched to the outcome.

Once the vector feature is mapped to the EO data, the researcher then needs to determine what value to assign. When exposure is represented by a gridded raster surface, the researcher must decide whether to preserve the value of each pixel or grid cell (e.g., rainfall quantity, night-lights radiance) or summarize those values to a different spatial support (Gotway and Young, 2002; Hengl, 2006). In the case of continuous raster surfaces (e.g., temperature, soil moisture), points can be assigned using the value of the containing cell, a nearest-cell rule, interpolation, or an average over a surrounding buffer. Lines can be assigned the value of the cell in which it crosses, or by a weighted value based on linear distance, travel time, or wind direction and speed. Polygons can be assigned area-weighted means, medians, totals, maximums, minimums, or shares above a threshold. In the case of categorical rasters (e.g., land-cover, crop-type maps), the relevant summary may be the modal class or the share of the polygon in each class rather than an arithmetic mean. The appropriate value or summary statistic depends on the outcome.

Regardless of whether exposure is mapped to a single cell, a set of cells, or a polygon, the researcher must decide how the assignment operates through time. Some treatments can be represented as a single cross-section: cells inside a protected area after designation, cells crossed by a newly opened road, or cells within a flood extent on a specific date. Others require a time series of rasters, vectors, or event layers (Kwan, 2012). Daily flood maps can be converted into the number of inundated days per cell; monthly NDVI anomalies can be summarized over crop-development stages; and conflict events can be counted within buffers by month, season, or year. If the mechanism is lagged or cumulative, each cell or unit may need an exposure history rather than a one-period value. The protocol should specify whether values retain their time-specific variation, are averaged over a window, accumulated, lagged, converted into a moving average, or collapsed to an indicator for any exposure during the relevant period.

The researcher also needs to determine whether the empirical variable captures treatment occurrence, treatment intensity, or a dose-response relationship (Hirano and Imbens, 2005). A rasterized treatment may assign all cells inside a policy boundary the same binary value, such as protected or not protected, electrified or not electrified, flooded or not flooded. Alternatively, treated cells may receive different values: distance to the nearest road, share of area protected, years since designation, number of inundated days, maximum flood depth, bombing density, smoke concentration, drought severity, or vegetation loss. These choices correspond to different estimands. A binary variable estimates the effect of crossing into treatment status. A categorical variable allows effects to vary across severity classes. A continuous variable estimates how outcomes change with marginal changes in exposure, assuming the scale is meaningful and comparable. For example, a flood event can be coded as any inundation, duration of inundation, maximum water depth, or share of a parcel inundated; a drought can be coded as any dry week, number of dry weeks, categorical drought severity, or a continuous index value. The empirical treatment variable should thus be defined before estimation as an indicator, count, rate, share, categorical class, or continuous surface.

Finally, the researcher must define how uncertainty in the mapping from conceptual treatment to empirical value will be handled. Uncertainty can enter through imprecise coordinates, displaced survey points, vector boundaries, raster classification error, cloud contamination, missing observations, temporal mismatch, projection choices, resampling methods, or the rule used to decide whether a cell is inside a polygon or crossed by a line. Some uncertainty can be addressed through data-quality filters, such as cloud masks, QA bands, or exclusion of low-confidence classifications (Elmes et al., 2020; Olofsson et al., 2014). Other uncertainty requires sensitivity analysis: comparing centroid-based, all-touched, and area-weighted rasterization rules; expanding or shrinking polygon boundaries; varying buffer distances or cone angles; simulating coordinate displacement; testing alternative temporal windows; or comparing multiple EO products that measure the same construct. In some cases, uncertainty can be retained directly in the treatment variable, such as using land-cover probabilities instead of hard classes or weighting cells by the probability of flood, fire, crop type, or exposure. Whatever approach is used, the researcher should document the coordinate reference system, resampling method, inclusion rule, threshold, quality filters, and robustness checks used to translate the conceptual definition of treatment into empirical values.

## 3.2. Assigning Exposure to a Unit of Analysis

Assigning treatment exposure to a unit of analysis is the step that makes the empirical treatment variable usable for causal analysis. After treatment has been represented as a point, line, polygon, raster surface, time series, or derived index, the researcher must decide how to attach that value to the unit for which outcomes are observed. That unit may be a parcel of land, a household or firm location, a village, administrative unit, protected area, road segment, or pixel. The assignment depends on three related features of the data: the spatial support of the EO product, the geometry of the socioeconomic observation, and the precision with which the socioeconomic unit's location is recorded. A spatial join is therefore not only a technical merge. It determines whether treatment is interpreted as exposure at a household location, exposure on cultivated land, exposure averaged over an administrative unit, or exposure over some other unit relevant to the estimand.

### 3.2.1. Collecting and Matching Your Own Data to EO Data

When researchers collect their own socioeconomic data, they have the greatest opportunity to align the unit of analysis with the unit of treatment. A critical decision then is what geometry needs to be georegistered? A household point may be sufficient when the outcome concerns household exposure to a weather surface. A field centroid or plot polygon may be necessary when the outcome concerns crop choice, yield, soil moisture, or land degradation. A school, clinic, irrigation canal, road segment, or conservation area may require a different geometry again. The typical method for capturing geolocation data, be it points, lines, or polygons, involves visiting the location with a GPS device or a GPS-enabled tablet or phone. Recent survey methods also allow enumerators or respondents to delineate fields using high-resolution imagery on tablets, which can reduce field time for dispersed or hard-to-reach plots (Masuda et al., 2020).

Once the socioeconomic unit has been georegistered, assigning EO values to that unit requires following the assignment protocol. Points can usually be matched to the grid cell in which they fall, to the nearest cell, or to a local average around the point. Polygons that fit within a grid cell, such as a field within a cell from a coarse weather project, can be treated in the same way a point is treated. For polygons that cover multiple grid cells, the researcher must choose how to summarize the values in the cells they cover (Gotway and Young, 2002; Hengl, 2006). A field polygon that intersects a grid cell can be assigned an area-weighted average. A district that intersects a soil map can be assigned the share of land of each soil type. A parcel that intersects a flood map can be assigned an indicator for any inundation. These choices are especially consequential when the socioeconomic geometry and EO resolution differ.

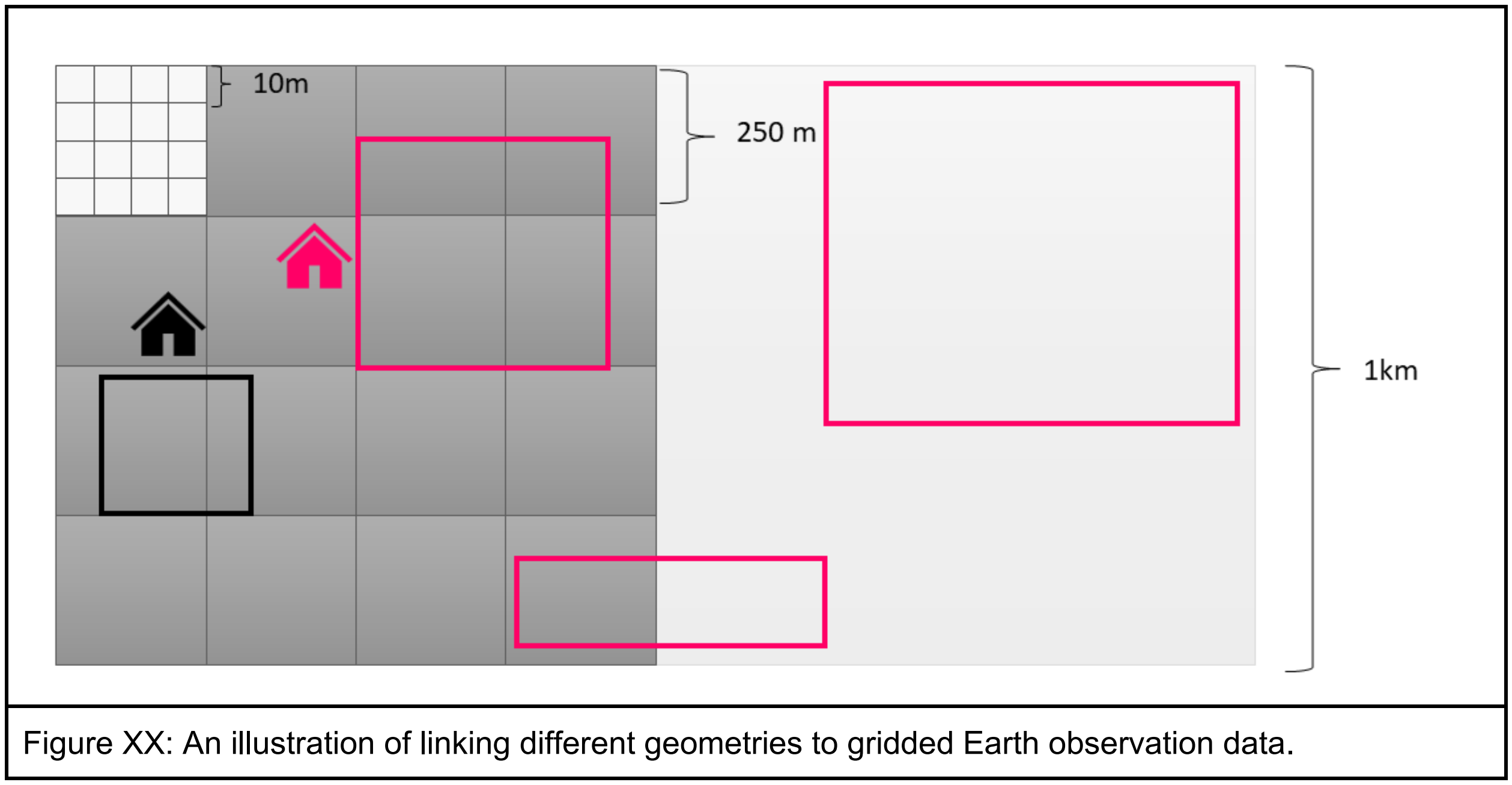

Figure XX: An illustration of linking different geometries to gridded Earth observation data.

Figure XX illustrates the challenges of linking household and field vector data (in the form of points or polygons) to high and moderate resolution EO data. There are two demonstrative examples of households, with their accompanying agricultural parcels. One household (in black) has a single parcel, directly adjacent to the house and roughly 250 $m^2$ in size. The second household (in red) has multiple plots, some of which are half a kilometer away from the house, and vary in size and shape. A household point is rarely located in the same high-resolution pixel as all of the household's fields, and a household may cultivate plots that are distant from the dwelling. In such cases, assigning field conditions from the household coordinate can create measurement error even when the coordinate itself is accurate.

The spatial support chosen for assignment also determines how much variation is preserved or lost. Moving from pixels to grid cells to fields, villages, administrative units, watersheds, or market areas can reduce noise and align the treatment with the scale at which outcomes are observed but it can also change the estimand. This is the practical relevance of the Modifiable Area Unit Problem (MAUP) (Openshaw and Taylor, 1979; Wong, 2004). The MAUP refers to the sensitivity of statistical results to the spatial aggregation unit. For example, aggregating crop yields to the administrative unit by calculating mean yields compared to aggregating to an agro-ecological zone using mean yields. Both approaches will produce a map of average yields, but the spatial patterns may differ due to the underlying aggregation. Neither support is inherently superior. The appropriate choice depends on whether the mechanism impacting crop yield operates through policy, markets, biophysical conditions, or some combination of these. Researchers should therefore choose aggregation units that match the causal process and, where possible, minimize meaningful within-unit heterogeneity (Garcia & Heilmayr, 2024). One may need to test alternative zoning assumptions to determine the most appropriate aggregation unit. A study examining MAUP effects on conservation programs found that both too small and too large aggregation units can each result in different forms of bias, based on the aggregation of data, and recommend that “researchers should prioritize the relevant scale for the dependent variable” (Avelino et al., 2016).

<< Box 1 starts here >>

**Box 1: Collecting Ground Reference Data**

If researchers are already in the field collecting socioeconomic data, this is an opportunity to also collect ground reference data for training, testing, calibration, debiasing, and independent validation of EO-derived measures. High- and moderate-resolution EO products used to measure crop type, on-field management, soil or water regimes, biomass, yield proxies, forest cover, or land degradation often improve substantially when paired with ground reference data (Deines et al., 2019a; Deines et al., 2019b; Jain, 2020). There is also growing awareness of the value of calibrating lower-resolution satellite weather data using ground-measured data from the study area (Josephson et al., 2026). When enumerators are already in the field, collecting information such as crop type, yield, area planted, rainfall, temperature, management practices, GPS locations can often be added at relatively low marginal cost.

Ground reference collection should be designed with the intended EO application in mind. Researchers face tradeoffs between dense measurement in a small area, which can improve local precision, and broader geographic coverage, which can improve representativeness across the treatment geometry. The number and distribution of reference observations should reflect the scope of the task, the heterogeneity of the phenomenon, and the modeling approach. Clear metadata are also essential: sampling frames, timing, geolocation accuracy, measurement protocols, enumerator training, missingness, and linkage rules should be documented so that ground data can be credibly matched to EO products and reused in validation. Training, test, and validation data should be kept conceptually and empirically distinct, especially when machine-learning models are used to generate EO-derived measures (Elmes et al., 2020).

However, one should not assume that ground reference data is error-free. Self-reported information is vulnerable to recall and social-desirability bias; crop cuts from sub-plot capture only part of field-level yield variation; and reference labels may contain sampling or classification error introduced during annotation. A further concern is representativeness: the locations sampled for ground reference collection may not reflect the broader population of fields, farms, firms, agro-ecological conditions, or the policy environment to which the analysis seeks to generalize, and this sampling limitation can compound other sources of error. For instance, field trials tend to represent unusually well-managed plots with attentive oversight, and measured outcomes such as yields may be systematically higher than those in surrounding areas, making them a potentially misleading benchmark for population-level inference.

The appropriate response is, first, to document likely sources of uncertainty explicitly and triangulate across multiple data types wherever possible (Elmes et al., 2020; Proctor et al., 2023; Alix-Garcia and Millimet, 2023). Where resources and context allow, researchers should prioritize independently collected, high-accuracy measurements (such as professionally supervised crop cuts or laboratory-validated samples), particularly when administrative or self-reported data are known to carry systematic biases relevant to the outcome of interest. Positional accuracy deserves separate attention: hand-held instruments and imprecise georeferencing can introduce spatial noise or systematic bias that affects how ground observations are matched to EO pixels, and georeferencing protocols should be documented alongside other metadata, including instrument type and its expected positional error. Where high-accuracy data are unavailable, simulating plausible measurement error and propagating it through the analysis can help bound its influence on estimates.

Public benchmark datasets such as PANGAEA and Fields of the World, as well as crowdsourced reference data, can be useful complements, but they do not remove the need to assess whether reference data represent the study area and treatment geometry of interest (Fritz et al., 2017; Marsocci et al., 2024; Kerner et al., 2025). When ground data are unavailable, researchers should be explicit about uncertainty, compare multiple EO products where possible, and consider ensemble, Bayesian measurement-error, or consensus-based classification approaches before labeling exposure. The key is to decide before fieldwork what geospatial data the project requires and how ground reference data will support matching, validation, calibration, or model training.

<< Box 1 ends here>>

### 3.2.2. Matching Public Use and/or Administrative Data to EO Data

Instead of relying on self-collected data, many projects rely on public-use surveys or administrative data. In those settings, the researcher has less control over geolocation or how the unit record is defined. In public-use data, precise household, firm, or village coordinates are unavailable due to ethical and legal concerns, particularly under regulations like the EU's General Data Protection Regulation (GDPR). Household or homestead coordinates are considered personally identifiable information (PII) and are thus protected by law or by the standards of informed consent (Josephson and Michler, 2024). To preserve privacy when publishing data, data administrators implement statistical disclosure limitation (SDL). SDL methods distort data, preserving privacy but reducing data accuracy and interoperability (Jolliffe et al., 2021). Interoperability relates to the ease with which different data sources can be joined or merged, including geographic coordinates or common geographic identifiers.

There are a variety of SDL methods to spatially anonymize GPS coordinates prior to the public release of data. One method is to simply not release individual-level coordinates but instead release coordinates for an aggregated area (i.e., cluster), like a country or enumeration area (EA). Figure YY provides a visualization of a number of different coordinate masking techniques and how these techniques affect which gridcell a coordinate would fall into. For example, the public-use GPS coordinates in USAID's Demographic and Health Survey (DHS) and the World Bank's LSMS-Integrated Surveys on Agriculture (LSMS-ISA) are cluster-based and randomly displaced anywhere between 0 and 2 km for urban clusters and between 0 and 5 km for rural clusters (Perez-Heydrich et al., 2016; Warren et al., 2015; Blankespoor et al., 2021).

Whether that loss of precision matters depends on the EO product and the research question. Spatial anonymization may have limited effect when matching locations to coarse, spatially smooth, weather products, because the grid cells themselves may be several kilometers wide and because nearby grid cells often have similar rainfall or temperature values (Michler et. al, 2022). The same displacement can be highly consequential when matching to high-resolution imagery, crop maps, flood extents, land-cover classifications, pollution gradients, or plot-level production. A displaced point can fall in the wrong plot, the wrong land-cover class, the wrong side of a flood boundary, or a different high-resolution pixel. SDL also reduces detectable within-area variation, as unit records in the public-use data are often assigned the same coordinate location as neighboring unit records. Researchers looking to match public-use socioeconomic data with EO data should therefore identify the SDL procedure, compare it with the spatial resolution and heterogeneity of the EO product, and avoid research questions that require finer location accuracy than the released data supports. Table XX in Further Resources provides a list of popular public-use data and the SDL methods used to anonymize locations.

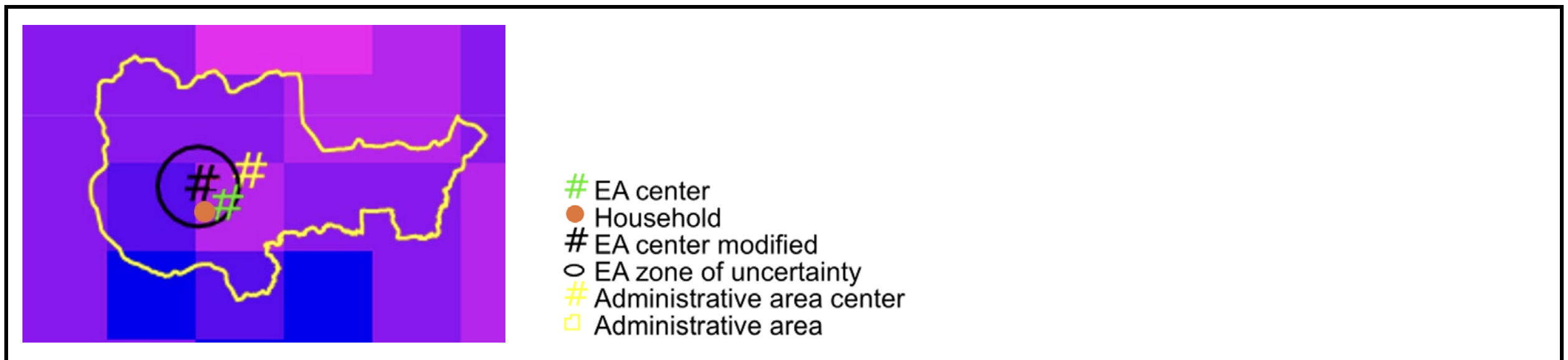

Figure YY: The figure presents the different anonymization methods and how the measurement of the anonymization method would vary across a particular gridded remote sensing precipitation product. The different colors indicate different measurements of rainfall, with darker blue indicating more rainfall and lighter pink indicating less rainfall. In this case: EA = Enumeration area.

Administrative data introduces a related problem. Outcomes may be reported for program jurisdictions, EAs, health-facility catchments, school districts, municipalities, or districts, while EO products are measured on grids or in polygons that do not align with those boundaries. Assigning exposure then requires a crosswalk: area weights, population weights, crop-area weights, land-cover weights, or another rule for mapping values from one support to another. Boundary vintages and coordinate reference systems should be checked carefully, since administrative boundaries may change over time and misaligned boundaries can create artificial changes in exposure.

Researchers should treat assignment to the unit of analysis as part of the treatment geometry protocol. The protocol should state the geometry of the socioeconomic unit, the EO product and resolution, the coordinate system, the spatial join or crosswalk rule, the aggregation function, and any assumptions about displacement, boundary uncertainty, or missing data. It should also explain why the assigned value corresponds to the mechanism being tested. Across these examples, the broader question is how to preserve the variation that matters for inference while avoiding exposure definitions that are too coarse, too noisy, or otherwise misaligned with outcome and estimand.

# 4. Design Choices for Identification

Treatment geometry is not only a measurement problem, but it is also an identification problem. The chosen geometry determines which units are treated, which units are controls, what variation identifies the effect, and what assumptions are required for causal interpretation. For that reason, treatment geometry should be treated as an explicit component of the research design rather than as a technical preprocessing step.

We distinguish between three related design choices. First, treatment geometry clarifies the estimand by defining the causal object being estimated. Second, it shapes comparison-group construction by determining which units are untreated, indirectly treated, or differentially exposed. Third, it creates or mitigates threats to identification, including spillovers, treatment misclassification, exposure measurement error, and specification search. We conclude this section with a discussion of sensitivity analysis that can strengthen arguments for causal identification.

## 4.1. Estimand Clarity

The first design choice is to align treatment geometry with the estimand. Different geometries answer different causal questions. A geometry that defines treatment only within a formal policy boundary is appropriate for estimating a direct effect inside that boundary, but not necessarily for estimating a total effect that includes leakage, displacement, or other spillovers. A binary treatment indicator estimates the effect of crossing into treatment status, while a continuous or categorical exposure measure estimates how outcomes change with intensity, duration, or severity. A spatially bounded treatment estimates a different quantity than one defined through distance, direction, networks, or cumulative exposure.

This distinction matters because treatment geometry defines the variation used to identify the causal effect. Aggregating forest-loss pixels to protected areas, rainfall grid cells to districts, pollution surfaces to buffers, or field-level vegetation indices to households changes not only the scale of measurement but also the interpretation of the coefficient. A district-level drought variable estimates the effect of district-average exposure, not field-level water stress. These choices raise the MAUP, as statistical relationships may change when the same underlying data are summarized using different spatial units or boundary systems (Openshaw and Taylor, 1979; Wong, 2004). Aggregation may reduce noise, but it may also smooth away the heterogeneity necessary for identification (Avelino et al., 2016; Blackman et al., 2024).

Estimand clarity also requires deciding how much realism to build into the treatment geometry. Simple geometries, such as buffers, administrative boundaries, or binary treated/untreated polygons, are transparent and feasible with limited data (Goodchild, 1992). More process-based geometries, such as hydrological networks, downwind exposure measures, travel-time catchments, or modeled pollution surfaces, may better approximate the exposure process but introduce additional assumptions and opportunities for specification search. The appropriate geometry is not necessarily the most complex one, but the one that best matches the estimand, causal pathway, data quality, and identifying variation.

## 4.2. Comparison Group Construction

The second design choice is to define credible comparison units. Treatment geometry determines which units are classified as treated, untreated, indirectly treated, or differentially exposed. These classifications shape the comparison being made and the assumptions needed for causal interpretation.

In spatial designs, nearby untreated units may be attractive controls because they may share similar geography, markets, institutions, and environmental conditions. But those same units may also be the most likely to experience spillovers. More distant controls may be less exposed to spillovers but less comparable. This trade-off is central to spatial causal inference: the best comparison units are not simply the nearest untreated units but the units that are plausibly untreated through the relevant causal pathway while remaining comparable on other dimensions (Keele and Titiunik, 2015).

Directional and network-based treatment sharpens the tradeoff. Upwind or upstream units may be useful comparisons because they are geographically close but less exposed through the pathway that transmits treatment. Orthogonal units, units outside a transport corridor, or units disconnected from an ownership, market, or supply-chain network may serve as comparison or placebo groups. These comparison groups are strongest when they are defined by the underlying mechanism before estimation rather than selected after observing results.

Spillovers should therefore be handled as part of comparison-group construction, not only as a post-estimation robustness concern. If untreated units may be affected by treatment, researchers should decide whether to exclude them from the comparison group, model them as indirectly treated, use them to estimate spillover effects, or incorporate them into the estimand. Each choice changes the interpretation of the estimate. Excluding likely spillover zones may help recover a direct effect, while including them may be necessary for estimating total effects. Modeling spillovers across distance bands, network links, or directional pathways can make the comparison more informative but only if the geometry corresponds to the mechanism being tested.

## 4.3. Threats to Identification

Treatment geometry can strengthen identification when it aligns exposure with the causal pathway, but it can also weaken identification when the geometry misclassifies treatment, contaminates comparisons, or embeds arbitrary assumptions. Three threats are especially common in GIE: interference, treatment misclassification, and nonclassical exposure measurement error.

The first threat is interference. Choices about the estimand and comparison group have direct implications for the Stable Unit Treatment Value Assumption (SUTVA). SUTVA is violated when control units are affected by treatment assignment elsewhere (Hudgens and Halloran, 2008; Aronow and Samii, 2017). This can occur through geographic proximity, market responses, mobility, hydrology or atmospheric transport, ecological processes or institutional networks. If nominally untreated units are indirectly exposed, estimated treatment effects may no longer recover the intended direct effect. Depending on the direction of spillovers, estimates may be attenuated, exaggerated, or reinterpreted as a mixture of direct and indirect effects.

The second threat is misclassification. Misclassification occurs when the empirical treatment variable assigns the wrong treatment status to a unit: a treated unit is coded as untreated, or an untreated unit is coded as treated. This can arise from imprecise coordinates, displaced survey points, uncertain vector boundaries, raster classification error, cloud contamination, or rules for deciding whether a cell is inside a polygon or crossed by a line. Misclassification is especially consequential near boundaries, thresholds, and narrow exposure pathways. The resulting bias depends on whether false positives and false negatives are random or correlated with treatment, outcomes, or determinants of treatment.

The third threat is exposure measurement error. Measurement error is broader than misclassification because a unit may be correctly classified as exposed but assigned the wrong intensity, duration, timing, or pathway of exposure. In geospatial settings, this error may be non-classical rather than classical random noise (Bound et al., 2001). For example, measurement error in satellite-derived forest-cover and deforestation data can be correlated with geography and characteristics of the raw data (Alix-García and Millimet, 2023). Similar concerns arise when cloud cover, slope, tree canopy structure, sensor resolution, detection thresholds, classification algorithms, or training-data quality affect whether and how exposure is observed. If measurement error is correlated with treatment status, baseline risk, enforcement, land cover, or outcomes, bias need not be simple attenuation. A null or small coefficient should therefore not automatically be interpreted as evidence of no effect.

In some designs, researchers may choose to sacrifice some descriptive accuracy in order to isolate exogenous variation. This is especially relevant for instrumental variables, fuzzy regression discontinuity, and related designs. For example, researchers may use relative location downwind from a pollution source or evapotranspiring forest as an exogenous source of variation rather than directly measuring realized pollution or rainfall exposure (Deryugina et al., 2019; Grosset-Touba et al., 2024). Similarly, a fuzzy regression discontinuity design may use an arbitrary or quasi-random feature of assignment rather than final treatment receipt if the latter is endogenous (Jordán and Heilmayr, 2024). In these cases, the geometry should be justified by the research design. What matters is not whether it perfectly describes realized exposure, but whether it isolates the variation needed for the target estimand.

## 4.4. Sensitivity Analysis and Placebo Tests

Because treatment geometry is part of the identification strategy, sensitivity analysis should be treated as a core diagnostic rather than as an afterthought (Kwan, 2012). Different treatment geometries will inevitably lead to different estimated outcomes but the variance in those outcomes should not be sensitive to small or arbitrary design choices. Therefore, researchers should examine whether results change under plausible alternative radii, time windows, thresholds, aggregation rules, directional assumptions, boundary definitions, and spatial supports. Spatial-support checks are especially important when treatment must be aggregated from pixels or grid cells to polygons, administrative units, watersheds, buffers, or market areas. Measurement-error concerns also imply testing alternative EO products, classification thresholds, confidence scores, quality filters, cloud masks, spatial displacement assumptions, and treatment assignment rules. Robustness to arbitrary alternatives is not sufficient to prove identification, but sensitivity to reasonable alternatives is informative about how much results depend on the chosen geometry.

Placebo and falsification tests provide complementary evidence. Researchers should test whether effects appear in places where the mechanism predicts they should not, such as upwind, upstream, pre-treatment, orthogonal, or otherwise unexposed units. They should also consider whether estimates are driven by units near uncertain boundaries, by locations with weak data support, by areas where remote sensing accuracy is known to be low, or by aggregation choices that obscure within-unit heterogeneity. These diagnostics help reveal whether the estimated effect is consistent with the hypothesized treatment pathway or instead reflects spillovers, misclassification, nonclassical measurement error, or dependence on a particular spatial support. The goal is to make the identifying assumptions embedded in the treatment geometry explicit, defensible, and empirically assessable.

# 5. Case Studies of Decisions and Challenges Facing Researchers

The case studies in this section illustrate how treatment geometry choices arise across diverse empirical settings. They also highlight a set of recurring spatial and temporal challenges that arise when combining EO data with socioeconomic data, be it survey or administrative, for GIE. We organize the discussion around a small number of concrete problems, emphasizing why each issue matters for inference, how researchers have addressed it, and what those choices imply for findings. Across the section, we focus on how spatial and temporal mismatches can shape treatment assignment, exposure measurement, spillovers, and interpretation in applied work (Jain, 2020; Wuepper et al., 2025). Table 1 syntheses these cases, in brief.

## Case Study 1: Spatial Spillovers

**The challenge**: In many contexts, treatment in one location can influence an outcome of interest in other locations. A common example is the case of forest conservation, where protected areas, certification programs, or land-use restrictions may alter behavior outside the formal treatment boundary (Pfaff and Robalino, 2017). The establishment of protected areas may increase deforestation in proximate locations (e.g. leakage) by shifting local commodity prices or displacing people. Alternatively, protected areas may reduce nearby deforestation if conservation norms, enforcement, tourism, or land-use expectations spill over into adjacent communities. In either case, the treatment geometry cannot be limited mechanically to the protected polygon or certified concession if the intervention changes outcomes beyond that boundary.

**Why it matters**: Understanding spillovers is important for two distinct reasons. First, spillovers can violate the SUTVA, a core assumption underpinning the causal interpretation of many research designs. If nominally untreated control units are affected by treatment through leakage, behavioral diffusion, price changes, or displacement, then common research designs will yield biased estimates of the direct treatment effect. Second, spillovers may be part of the policy effect of interest. If a protected area merely moves deforestation outside of its borders rather than reducing total deforestation, the environmental benefits of protection may be much smaller than estimated from treated polygons alone.

**Example - Forest conservation**: Forest conservation provides a clear example of why potential spillovers should be one element addressed by treatment geometry. For example, Robalino, Pfaff and Villalobos (2017) show that spillovers from protected areas in Costa Rica are heterogeneous in ways that align with theory: spillovers differ with proximity to roads and with distance from park entrances where eco-tourism values are high. These patterns suggest that spillovers are not simply a function of distance from a protected boundary but of the mechanisms through which land-use incentives are transmitted.

Heilmayr, Carlson, and Benedict (2020) document an example of spillovers operating through economic relationships rather than local proximity. In their case, palm oil corporations that adopt sustainability certificates on some of their concessions shift deforestation behavior onto their non-certified concessions, including concessions far outside of local markets for palm fruit bunches.

**Practical steps for analysis**: When estimating spillovers directly, or to protect the validity of SUTVA, researchers should begin by building a theoretical model of the spatio-temporal patterns of spillovers. In the simplest case, researchers can compare post-treatment changes in outcomes across untreated units based on their distance to treated units using rings or buffers around treated areas to test whether outcomes change with proximity (Andam et al., 2008). This approach is transparent and useful but it assumes that spillovers decay with distance.

More sophisticated analyses define spillover exposure using the mechanism expected to transmit the effect. If leakage operates through local land markets, relevant comparison units may be nearby parcels, communities, or jurisdictions. If spillovers operate through roads, market access, or park entrances, the treatment geometry should incorporate those features. If spillovers operate through firms, ownership networks, or supply chains, then researchers may need to link spatial units by economic relationship rather than by distance alone.

The empirical workflow should thus make spillover assignment explicit. Researchers should map the directly treated units, identify the untreated units that could plausibly be indirectly affected, and decide whether those units are excluded from the comparison group, coded as indirectly treated, or used to estimate spillover effects directly. They should then test sensitivity to alternative spillover geometries, including different buffer widths, road- or market-access measures, temporal windows, and definitions of connected units. Placebo comparisons can be useful when theory predicts that some nearby units should not be affected, or that spillovers should appear only after treatment begins. The goal is not only to avoid contaminated controls, but to make the assumed spatial and temporal reach of spillovers visible, reproducible, and aligned with the mechanism being tested.

**Key risks and assumptions**: The main risk is misclassifying indirectly affected units as untreated. If control units are exposed to spillovers, estimated treatment effects may be attenuated, exaggerated, or otherwise biased, depending on the direction of the spillover. Leakage that increases deforestation in nearby controls can make protected areas appear more effective than they are within the broader landscape, while beneficial spillovers that reduce deforestation in nearby controls can make protected areas appear less effective than they are. A second risk is assuming that spillovers are purely spatial when they may instead operate through economic, institutional, or ownership networks. Distance-based buffers are useful, but they may miss spillovers transmitted through commodity markets, firms, migrant networks, enforcement behavior, or conservation norms.

**What to remember**: Spatial spillovers mean that the boundary of treatment is not always the boundary of the policy. The forest conservation literature highlights several broad lessons for defining treatments when spillovers exist: first, spillovers can change outcomes outside of the formal area of treatment; second, ignoring spillovers can bias treatment effect estimates and the distort interpretation of policy effectiveness; and third, generating credible estimates of spillovers requires a deep understanding of the underlying mechanisms through which treatment affects other units.

## Case Study 2: Directionality and Transport

**The challenge:** EO data is frequently used to evaluate exposure to environmental hazards. These range from pollution of various types (e.g., air or water) to climate hazards. EO data is especially useful in this context because pollutants frequently come from specific sources that can be identified and mapped using EO data. Often it is easier to identify and map the location of specific polluting sources than to identify and map high frequency measures of the pollution itself. Taking this approach requires a choice about how to assign treatment.

**Why it matters**: Early work that used EO data to assign treatment by pollution often used circular buffers around polluting sources (sometimes with a donut to exclude places especially close to sources) to define treatment. But that approach ignores the dynamics of pollution dispersion in ways that can bias estimates by assigning equal treatment status to places that, in reality, receive very different doses (Sullivan, 2016).

**Example - Air pollution**: In the case of air pollution, the most sophisticated method to assign treatment would be to run an atmospheric circulation model that considers the identified source location and calculates plume dispersion based on complex atmospheric chemistry, physics, and meteorological conditions. This approach is beyond the scope of most projects concerned with socio-economic impacts, although reduced complexity models like AP4 (Muller and Mendelsohn, 2007) and InMap (Tessum et al., 2017) are rapidly making this approach more feasible.

When full dispersion modeling is not feasible, researchers often use wind-based treatment assignment (Miller et al., 2019; Zabrocki et al., 2022). The approach is to identify the emitting source, measure wind direction and speed over the relevant time window, and classify units as treated when they are downwind of the source. This turns a point-source treatment into a directional exposure geometry. The treatment variable can then be defined as an indicator for ever being downwind, the number of downwind days, the share of days downwind during a month or year, or a weighted measure that combines downwind exposure with distance from the source. This leaves two core design choices: how to aggregate changing wind patterns over time, and how to define the downwind exposure area in space. Those choices should follow the pollutant, source type, outcome, and temporal window of interest rather than being treated as purely technical parameters.

**Practical steps for analysis**: There are no hard and fast rules to guide these choices. For the choice of aggregation, the most common approach is to count days when a potentially treated unit is downwind of a pollution source and define treatment based on either the raw number of days or the share of a year (or month or other relevant aggregation period) that it is downwind.

That leaves the determination of what is downwind. The common approach here is to define a cone of various angles with the vertex at the emitting source and a variety of heights. The choice of angle and height can be made based on the type of emission. For example, power plants emitting from tall stacks might justify a narrow angle and greater height than surface level agricultural burning. Or, based on wind speed: higher wind speed might again suggest narrow angles and greater heights. It is good practice to show robustness to a variety of choices, regardless of the primary cone used.

It can also be useful to define specific sets of non-treated units based on whether they are upwind of a pollution source and/or orthogonal to the direction of the wind. Depending on the question being tested these units can serve as useful placebo units.

**Key risks and assumptions**: The main assumption is that the wind is the primary determinant of how pollution is spread from an emitting source, and so using wind direction and speed serve as a good proxy for how that pollution spreads. But atmospheric transport is complex and the risk is that there are more complicated processes that determine who is or is not exposed to pollution. The extent to which this risk is meaningful will depend on the location of the study and the types of pollution being studied, as wind is an important determinant in the transport of some pollutants more than others. It is worth noting that proxies for how pollution travels rarely work uniformly well in every geography.

**What to remember:** The main thing to remember is that pollution does not disperse randomly or uniformly from an emitting source. There are specific physical processes that dictate who will or will not be exposed to pollution from a given source. Those processes can be modelled in exacting detail but this is often computationally intensive. They can also be proxied using approximations of the forces that determine the spread of pollution. However, it is important to test a variety of different proxies and assess which is most appropriate for a given pollutant and context. It is also generally good practice to show that results are insensitive to reasonable changes in how the proxies are measured.

## Case Study 3: Sparse and Mismatched Data Inputs

**The challenge:** In GIE, researchers frequently face situations where the phenomenon of interest varies across space, but available datasets are sparse, unevenly distributed, or misaligned across spatial units. Exposure data are often collected at points (e.g., weather stations), while outcomes are measured over aggregated units (e.g., like household or district). Because the spatial supports of these datasets differ, analysts must decide how to transform, interpolate, or aggregate data to make them compatible for analysis.

**Why it matters**: Transforming sparse or mismatched data introduces implicit assumptions that can shape results. Interpolating point measurements into a continuous surface assumes spatial smoothness that may not exist (Li and Heap, 2014). Aggregating to administrative units can smooth away meaningful variation, especially for directional or network-driven processes such as smoke transport, river flow, or groundwater depletion. Recognizing these assumptions is essential for producing credible impact estimates.

**Example - Environmental exposure**: Suppose we aim to estimate the effect of an environmental stressor (e.g., drought, wildfire smoke, or aquifer depletion) on households or assets within a defined treatment geometry. Exposure may be observed at a limited set of monitoring points (e.g., air pollution sensors), while outcomes are measured over broader administrative units. A typical data analysis workflow involves:

1. Interpolating point observations into a continuous surface via smoothing/filler algorithms (e.g., Savitzky-Golay, harmonic trend) (Cressie, 1993; Li and Heap, 2014).

2. Aggregating the surface to match the outcome units.

This workflow imposes assumptions about within-unit homogeneity and may misrepresent exposure for processes that are directional or network-based. Researchers should explicitly consider these assumptions before proceeding.

**Practical steps for analysis**: Before implementing any workflow designed to interpolate missing data, one should ask "Do my sensor/ground-reference data locations reasonably represent the spatial variation I am trying to measure?" If the answer is "no," then no amount of interpolation will produce data that either is accurate or has the necessary variation for causal identification.

However, if the answer is "yes," then the researcher must assess the spatial coverage in the relevant area. This can involve computing sensor density, distance to the nearest observation, and share of area within a given radius of sensors. Next, would be implementing the smoothing algorithm to fill in the data gaps. Once that is done, the researcher should test the sensitivity of the results by recomputing exposures using alternative aggregation schemes, such as pixel-level, buffer, or polygon-based approaches, as well as validating the existing interpolated map by withholding a subset of data before interpolating data. Finally, one should triangulate datasets by comparing multiple exposure products (e.g., alternative air pollution or smoke surfaces) and use disagreements between products to inform uncertainty.

**Key risks and assumptions**: Sparse and mismatched data pose several risks. First, interpolated surfaces are modeled estimates, not direct observations. This can give rise to the generated regressor problem in which standard errors in the main regression of interest are too small because they treat the modelled data as fixed/observed rather than estimated or generated (Pagan, 1984). Second, monitoring locations are often non-random: meteorological stations may be sited for forecasting purposes, and agricultural sensors are typically placed in controlled, uniform plots. This can introduce systematic bias, meaning that ground-based observations may not represent broader conditions and may systematically overestimate or underestimate a variable's true value in certain regions or conditions. Researchers should therefore treat "ground truth" as one of several possible estimates, not an absolute standard.

**What to remember:** When working with sparse or mismatched data, treatment geometry is partly constructed rather than directly observed. Transparent documentation of assumptions, combined with sensitivity analysis and triangulation, is essential for producing credible and interpretable estimates. Optional enhancements include a visual schematic of the translation from points to surface to aggregated units, a "common mistake" callout (e.g., treating interpolated values as ground truth), or code references for interpolation workflows.

## Case Study 4: Treatment Timing in the Presence of Heterogeneity

**The Challenge**: In many contexts, the effect of a treatment or shock depends not only on where it occurs, but also on when it occurs, relative to seasonal or management conditions. This is especially true in agricultural settings, where the same shock can have very different implications, depending on when it occurs during a plant's growth stage: just after seeding, at flowering, during fruit formation, at harvest, or outside the production season altogether. Agriculture production is often subject to many types of natural disturbances during the growing season: from weather shocks, such as droughts, flooding, and storm events, to pest and disease outbreaks. The timing of these events in relation to the growth stage is a key factor in the magnitude of impact of production. Even clearly observable shocks can have very different economic and production consequences depending on when they occur relative to those stages (Ortiz-Bobea, 2011). For example, it has been shown that rapeseed in parts of Northern Germany is affected more by droughts during flowering (Schmitt et al., 2022).

**Why it matters**: Timing matters because temporal misalignment can create measurement error in treatment assignment. If a researcher defines exposure to a shock using an annual average or a broad growing-season indicator, then the treatment variable may combine periods that differ sharply in relevance for production and household decision making (Lui et al., 2022; Ahvo et al., 2023). Timing also matters for interpretation: exposure during flowering answers a different question than exposure during the full agricultural year.

**Example - Agricultural shocks during the growing season**:

A locust swarm observed prior to planting may suggest little immediate crop loss, but the same swarm during other phases of crop production may cause significant damage (Chatterjee, 2020).

Similarly, tropical cyclones have differentiated impacts on crop production depending on when they make landfall and at what stage the crop is (Kunze, 2021; Bundy et al., 2023). A cyclone before seeding may affect land preparation, planting decisions, or expectations for yield, while a cyclone near harvest may reduce yields, destroy standing crops, disrupt harvest, and/or eradicate market access (Sivakumar, 2018; Yuen et al., 2022). Crop calendars, like those maintained by the Food and Agriculture Organization (FAO) of the United Nations, offer a practical framework for defining treatment timing because they identify when management decisions and phenological stages occur in specific places and for specific crops.

**Practical steps for analysis**: Researchers should begin by defining the outcome and the causal pathway linking exposure to that outcome. If the outcome is area planted, exposure during land preparation, tillage, or seeding may be the most relevant. If the outcome is yield conditional on planting, exposure during establishment, flowering, grain filling, or fruit formation may matter more. If the outcome is income, food insecurity, or asset sales, the relevant temporal footprint may include both the shock and the household response period, including labor reallocation, crop switching, or crop strategies.

Researchers should then align EO or hazard data to locally appropriate crop calendars and phenology, constructing stage-specific exposure measures. This can involve assigning each shock to a crop stage, calculating exposure within each stage, or creating indicators for whether an event occurred during particularly vulnerable windows. Where crop phenology varies across space or year, researchers should avoid relying only on national crop calendars and should incorporate local calendars, phenological indicators, or remotely sensed crop-stage measures where possible (Gong et al., 2024). Researchers should also test sensitivity to alternative temporal windows, especially when crop-stage timing or behavioral response are uncertain.

**Key risks and assumptions**: The main risk is that temporal aggregation masks the mechanism being estimated. Annual exposure measures can combine shocks that affect planting decision, crop growth, harvest outcomes, and post-harvest coping into a single variable, making estimates difficult to interpret. A second risk is assuming that crop calendars are fixed, when planting dates and crop stages vary by region, crop variety, irrigation, elevation, farmer behavior, and year-specific weather like El Niño and La Niña. A third risk is aligning exposure to data availability rather than to casual process, which can produce apparently precise but substantively weak treatment definitions.

**What to remember**: Treatment timing is part of treatment geometry. The consequences of exposure depend on when it occurs relative to management decisions, crop development, and seasonal cycles, not simply on whether it occurs within a calendar year. Temporally misaligned measures can bias treatment effect estimates and obfuscate mechanisms by combining exposure windows that operate on different margins. Credible temporal treatments definitions require a clear understanding of the process linking exposure to outcome and should be justified, documented, and tested against plausible alternatives.

## Case Study 5: Focusing on the Extensive or Intensive Margin

**The Challenge**: Once spatially explicit event data are combined with socioeconomic data, researchers must decide whether exposure should be defined as a discrete indicator of occurrence or as a continuous measure of intensity. A discrete treatment captures exposure at the extensive margin: whether an event occurred within the spatial and temporal window used to merge the EO and socioeconomic data. A continuous or categorical treatment captures exposure at the intensive margin: how severe, frequent, prolonged, or cumulative was exposure. This choice is not merely a modeling detail. It determines what variation identifies the effect and what causal statement the analysis can support.

**Why it matters**: Extensive- and intensive-margin treatments answer different questions. A binary variable estimates the effect of being exposed and is appropriate when occurrence itself is expected to trigger a response. A continuous or categorical variable estimates how outcomes change with greater exposure and is appropriate when cumulative damage, repeated exposure, or event severity is central to the mechanism. The decision depends on data availability, measurement accuracy, and whether the intensity measure has a meaningful and comparable scale. If intensity is poorly measured, a binary indicator may be more defensible. If intensity is well measured and substantively important, dichotomizing exposure can discard useful variation.

**Examples - Drought and conflict**: Drought exposure illustrates how one underlying variable can be represented in several ways. The standardized precipitation evapotranspiration index (SPEI) is widely used to measure unusually dry conditions because it incorporates precipitation and temperature through potential evapotranspiration and can be calculated across different time frames and geographic scales (Vincente-Serrano et al., 2010). SPEI is continuous, with positive values indicating water surplus and negative values indicating water deficit. When using the SPEI, researchers need to make decisions on the temporal aggregation period, the geographic unit, and whether to use SPEI continuously or convert it into categories or indicators.

For example, studies that are investigating effects of drought within a growing season have taken two approaches: calculating the SPEI with an aggregation time equivalent to the length of the growing season, such as 3 or 6 months (Antonelli et al., 2022; Meyer 2023; Murken et al., 2024; Pignède, 2025), or calculating the SPEI with a one-month aggregation (Uexkull et al., 2016). For example, Uexkull et al. (2016) studies the effects of droughts during the agricultural growing season on conflict events. They use monthly aggregation times for the SPEI and look at the fraction of “dry months” within the growing season, designating a month as dry if the SPEI is less than -1. The choice of threshold value is often based on the typically used and cited categorization from McKee et al., (1993), although this threshold and categorization is arbitrary. Another option would be to use probability distributions of the SPEI to determine classes (such as done by Pignède, 2025). Alternatively, other analyses have carried out sensitivity analysis to determine the best threshold (Antonelli et al., 2022).

Armed conflict provides a parallel example. Bloem et al. (2025) study herder-related violence and labor allocation in rural Nigerian households using a binary treatment variable for whether a violent event occurred within a specified distance of a household. Although fatalities could support a continuous intensity measure, the extensive-margin definition follows the idea that a single event can create expectations of violence sufficient to alter behavior (Vesco et al. 2025). By contrast, Appau et al. (2021) study the long-term impact of the Vietnam War on agricultural productivity using bombing intensity as a continuous treatment, because severity is central to the mechanism.

**Practical steps for analysis**: Researchers should begin by specifying whether the mechanism depends on occurrence, severity, frequency, duration, or accumulation. If behavior changes once any event occurs, an extensive-margin measure may be appropriate. If outcomes vary with magnitude or cumulative exposure, an intensive-margin measure is likely needed. Researchers should then assess whether the intensity measure is accurate and comparable across units. In many research settings, estimating both the intensive- and extensive-margin has relevance for policy.

When discretizing continuous measures, researchers should justify thresholds rather than treating conventional cutoffs as automatic. They should report whether results are sensitive to alternative aggregation periods, spatial units, thresholds, and functional forms. Where theory suggests nonlinear effects but not a precise cutoff, categories, bins, or multiple thresholds may be preferable to a single binary indicator.

**Key risks and assumptions**: The main risk with binary treatment is that it treats all exposed units as equivalent. A district with one dry month and a district with an entire dry season can receive the same treatment status if both cross the threshold. The main risk with a continuous treatment is false precision: the model may imply that marginal changes in a statistical index, fatality count, or bombing measure have stable and comparable meaning across places, periods, and populations. Both approaches require assumptions about whether measured exposure captures the mechanism of interest.

**What to remember**: The extensive/intensive-margin choice is part of treatment geometry. Discrete treatments identify the effect of being exposed versus not exposed; continuous treatments identify how outcomes change with greater exposure. Neither is inherently preferable. The right choice depends on mechanism, data quality, scale of measurement, and estimand. When reasonable alternatives exist, researchers should compare binary, categorical, and continuous definitions and document how those choices affect interpretation.

## Case Study 6: Georegistration of Ground Data

**The challenge**: When ground data, such as household surveys, agricultural plot measurements, enterprise surveys, or field validation data, are integrated with EO layers, location becomes both a critical input and a potential source of error (Michler et al., 2022). Small geolocation errors can change the EO covariates assigned to households, enterprises, agricultural plots, or other units of interest. This can affect treatment assignment, outcome measurement, targeting, or the construction of training and validation data for remote-sensing models. The relevant question is whether the level and type of error are consequential for the research design.

**Why it matters**: There is no universally applicable threshold for geolocation error. The tolerable level of error depends on the intended use of the data and the spatial scale of the analysis. In plot-level applications, such as crop mapping, yield estimation, or field-level treatment assignment, even small geolocation errors can shift a plot into a different pixel, crop class, or land-cover type. This can reduce the performance of remote sensing-based models and introduce measurement error into treatment and outcome variables (Perez-Heydrich et al., 2016; Warren et al., 2015; Michler et al., 2022). For analyses using coarser or smoother variables, such as gridded rainfall or temperature data, the same level of geolocation error may be less consequential because nearby pixels are more similar and grid cells are larger. The implication is that georegistration accuracy must be evaluated relative to the EO product, the spatial heterogeneity of the variable being measured, and the unit of analysis.

**Example - Crop mapping**: Agricultural plot data illustrate why the type of geometry collected matters. Researchers may record a single point, a plot centroid, a set of points on the boundary, or a line tracing the boundary. These choices are not interchangeable. For example, Azzari et al. (2021) show that satellite-based crop mapping models perform better when trained on ground data that include either full plot outlines or coordinates from all plot corners. Model performance falls when training data rely on a single point, although point centroid performs better than a point collected at a single plot corner. In their application, the choice of input geometry produced policy-relevant differences in the estimated area under maize cultivation, with suboptimal georegistration strategies overestimating the area under maize by up to 24%.

**Practical steps for analysis**: Researchers should begin by defining what geometry is needed for the intended EO application. A household point may be adequate when assigning coarse weather data to a household record. A field centroid may be sufficient for some plot-level summaries if fields are large relative to the EO grid and internally homogenous. Full plot polygons or multiple boundary points may be necessary for some applications, such as crop area or yield estimation, or when summarizing high-resolution EO data over cultivated land.

Researchers should then choose a georegistration workflow that balances accuracy, cost, time, and data management requirements. Handheld GPS devices generally provide strong location accuracy and are often preferred for collecting polygons, but performance can be hindered under dense tree canopy, near tall buildings, or in rugged terrain (Carletto et al., 2016). They also introduce data-management challenges: plot outlines stored as .gpx files must be linked correctly to the unique identifier in the survey data, and manual naming of tracks can result in high mismatch rates. GPS-enabled tablets used in Computer Assisted Personal Interview survey software (CAPI) reduce transcription errors and allow for real-time quality check, but generally offer lower positional accuracy than standalone GPS devices (Petruccelli et al., 2026). Delineation on georeferenced satellite imagery can reduce field time and is useful for distant, dangerous, or hard-to-access locations but performance depends on image quality, seasonal relevance, cloud cover, respondent familiarity with aerial imagery, and enumerator training, and landmarks to help identify the correct plot (Masuda et al., 2020; Dickinson and Gourlay, 2026).

**Key risks and assumptions**: The main risk is that ground data are treated as exact when they are themselves measured with error. A point may not represent the relevant unit if the household cultivated distant plots, if the point is collected at a plot corner rather than a centroid, or if the outcome depends on within-field variation. Polygon data may also contain errors if boundaries are simplified, if enumerators walk the wrong perimeter, or if satellite-image delineation misidentifies neighboring fields. A second risk is linkage error: accurate GPS tracks are not useful if they cannot be reliably matched to the correct household, enterprise, or plot record. A third risk is false precision. High-resolution EO products make geolocation error more consequential, not less, because small coordinate errors can change the assigned pixel value.

**What to remember**: Georegistration is part of treatment geometry because it determines how ground observations are linked to EO-derived treatment or outcome measures. The appropriate geometry and collection tool depend on the estimand, the spatial resolution of the EO product, the heterogeneity of the landscape, and the intended use of the data. Researchers should decide before fieldwork whether they need points, centroids, corners, polygons, or remotely delineated boundaries. They should also document how locations are collected and linked to survey records and, as always, test whether results are sensitive to plausible geolocation errors.

| Table 1: Synthesis of Treatment Geometry Choices across Cases | | | |
|---|---|---|---|
| Case | What must be defined | What researchers should ask | Implication for analysis |
| (1) Spatial spillovers | The spatial reach of direct and indirect treatment. | Can treated units affect outcomes in nominally untreated locations? Through what mechanism? | Treatment assignment may need to include indirect exposure, exclusion zones, or explicit spillover estimates. |
| (2) Directionality and transport | The path through which exposure moves across space. | Does exposure travel uniformly, or is it shaped by wind, water, terrain, roads, or other transport processes? | Distance-based buffers may need to be replaced by directional or process-based geometries. |
| (3) Sparse and mismatched data inputs | The spatial support on which exposure is measured and assigned. | Are point observations, pixels, and outcome units spatially compatible? | Interpolation and aggregation choices should be documented, validated, and tested for sensitivity. |
| (4) Treatment timing under heterogeneity | The relevant temporal window of exposure. | When does exposure matter for the outcome: before, during, or after the event or biological process? | Treatment measures should align with crop calendars, hazard timing, or behavioral response periods. |
| (5) Extensive versus intensive margin | The margin of exposure that identifies the causal effect. | Does the mechanism depend on occurrence, severity, frequency, duration, or accumulation? | Researchers should justify binary, categorical, or continuous exposure definitions and test alternatives. |
| (6) Georegistration of ground data | The location and shape of the ground unit linked to EO data. | Is a point, centroid, boundary, or polygon needed for the intended EO application? | Geolocation and linkage error should be treated as part of the measurement problem, not as incidental noise. |

# 6. Good and Better Practices: A Treatment Geometry Protocol

Viewed together, the six case studies suggest a common series of principles for defining treatment geometry. The researcher should begin by articulating the causal pathway connecting treatment to outcome. Next, the researcher should identify the spatial and temporal units over which that pathway operates. Then, the researcher should choose the EO and socioeconomic data products whose support, resolution, timing, and uncertainty are compatible with that pathway. The treatment variable should be constructed only after these choices are made. Finally, the researcher should evaluate whether alternative plausible geometries change the results or their interpretation.

This approach is particularly important because treatment geometry choices often become embedded in data-processing code before they are fully theorized. Buffer distances, raster extraction methods, aggregation windows, and spatial merges may appear as technical parameters but they encode assumptions about exposure and causality. A defensible analysis should make those assumptions visible. Maps, schematics, and descriptive diagnostics are useful tools for doing so but are not substitutes for a clear delineation and justification of the treatment geometry in the text of the paper or report. Researchers should state where treated and control units are located, how exposure varies across space and time, where data support is sparse, and which units lie near uncertain boundaries.

The goal is not to eliminate all uncertainty. In many geospatial applications, the true exposure process is only partially observed. Thus, the goal is to make the treatment geometry credible, transparent, and appropriately matched to the estimand. Paraphrasing Laitin (2013), a guiding principle in GIE should be to document when in the research process, and for what reason, a particular design choice was made. With that in mind, the following questions offer a practical starting point for treatment-geometry decisions before estimation begins. Not every question will be relevant to every setting, as which ones matter depends on the exposure type, the source geometry, and the causal pathway. The value is in working through them to identify which decisions are important in your context and ensuring those are made explicitly rather than by default.

**What do you want to learn?**

1. **What is the estimand?** Specify whether the analysis aims to estimate a direct effect within the treated area, a total effect including spillovers, an average effect of exposure, a marginal dose-response relationship, or another quantity.

2. **What is the causal pathway?** Describe how treatment is expected to affect the outcome. Identify whether the pathway is bounded, radial, directional, network-based, cumulative, seasonal, or mediated by markets, behavior, or institutions.

**How is treatment constructed?**

3. **What is the source geometry?** Define whether treatment originates from points, polygons, lines, networks, rasters, events, or continuous surfaces. Document the source dataset and any known measurement error.

4. **What are the "treatment" units?** Specify the units to which exposure is assigned: households, plots, firms, parcels, villages, administrative units, pixels, or another unit. Explain why this unit matches the outcome and mechanism.

5. **What spatial assignment rule is used?** State whether exposure is assigned by containment, intersection, distance, buffer, area-weighted overlap, population-weighted overlap, network distance, downwind location, downstream position, raster extraction, or another method.

6. **What temporal window is used?** Define treatment onset, duration, lags, accumulation, and reversibility. Justify whether the relevant window is daily, monthly, seasonal, annual, crop-stage-specific, policy-stage-specific, or cumulative.

7. **Is treatment discrete or continuous?** Explain whether the treatment captures the extensive margin, the intensive margin, or both. Justify any thresholds, bins, severity categories, or transformations.

**What could go wrong? Threats to validity/robustness**

8. **How are spillovers handled?** Identify whether treatment may affect nearby, connected, downwind, downstream, or market-linked units. State whether these units are excluded, modeled directly, used as comparison groups, or incorporated into the estimand.

9. **How is measurement error addressed?** Discuss georegistration error, boundary uncertainty, classification error, spatial anonymization, sensor sparsity, interpolation error, and mismatch between EO and socioeconomic data.

10. **How sensitive are results to alternative geometries?** Report robustness to plausible changes in buffer size, cone angle, temporal window, aggregation scale, exposure threshold, interpolation method, crosswalk rule, and treatment-control definitions.

**Can we reproduce it?**

11. **Can the geometry be visualized and/or reproduced?** Provide maps, code, metadata, coordinate reference systems, merge rules, and versioned data-processing steps. A reader should be able to understand not only the final treatment variable, but how and why it was constructed.

Box XX envisions how these questions and others fit into the four analytical design components discussed at the beginning of the chapter: (1) the source of treatment; (2) the unit that is treated; (3) the pathway of treatment; and (4) the temporal window of treatment. The result is a detailed protocol for implementing treatment geometry and turning it into a documented, defensible set of research-design decisions that readers can appropriately interpret.

In GIE, geometry is not a GIS detail; it is the empirical form of the causal question. The boundaries, buffers, time windows, assignment rules, and aggregation choices that define treatment also define the estimand, the comparison group, and the threats to identification. There is rarely a single correct geometry, but there are more and less defensible ones. The task is therefore not to eliminate uncertainty, but to make the assumed exposure process visible, testable, and reproducible. When researchers define treatment geometry explicitly, justify it theoretically, stress-test it empirically, and document it transparently, EO-derived measures can do more than fill data gaps: they can strengthen the credibility of causal inference.

| **Box 1: A Protocol for Defining Treatment Geometry in Geospatial Impact Evaluations** | |
|---|---|
| Use this protocol before estimation to make treatment geometry choices explicit. The goal is to clarify the estimand, construct credible comparison groups, identify threats to validity, and ensure that treatment assignment can be reproduced. | |
| **Elements** | **Guiding questions and suggested checks** |
| ***Define the Source of Treatment*** | |
| | *Estimand clarity*<br>● What is the treatment or exposure?<br>● What is the estimand (direct effect, total effect including spillovers, dose-response)?<br>● What causal path links treatment to outcome? |
| | *Comparison-group construction*<br>● Which units are untreated or differentially exposed?<br>● What variation identifies the effect? |
| | *Threats to identification*<br>● What source geometry represents treatment?<br>● What is the source dataset?<br>● What is the known measurement error? |
| ***Define the Unit of Treatment*** | |
| | *Estimand clarity*<br>● What units receive treatment (households, plots, pixels, admin units, etc.)?<br>● Why does the unit match the outcome, mechanism, and estimand? |
| | *Comparison-group construction*<br>● Which units remain eligible as comparisons after treatment is assigned?<br>● Are comparison and treatment units comparable in scale, geography, timing, and exposure opportunity? |
| | *Threats to identification*<br>● How could the chosen unit misclassify exposure or alter interpretation? |
| ***Define the Pathway of Treatment*** | |
| | *Estimand clarity*<br>● How does the treatment move through space or through systems? |
| | *Comparison-group construction*<br>● What units are credible given the pathway? |
| | *Threats to identification*<br>● How are spillover units handled (excluded, modeled, used as comparisons, or included in the estimand)? |
| ***Define the Temporal Window of Treatment*** | |
| | *Estimand clarity*<br>● What exposure window is relevant to the outcome? |
| | *Comparison-group construction*<br>● Which comparison units are temporally aligned and untreated during the relevant exposure period? |

|  | *Threats to identification*<br>● How could timing choices misclassify treatment or obscure mechanisms? |
|---|---|

## Acknowledgements

Generative artificial intelligence tools were used during the manuscript preparation process. Specifically, OpenAI's ChatGPT 5.5 Thinking was used to edit and format the article's text. The authors wrote, reviewed, and are responsible for the final content of the article.

## Contributors’ Biographies

Jeffrey D. Michler is Associate Professor of Agricultural and Resource Economics at the University of Arizona and Co-Director of the AIDE Lab. His research applies industrial organization, contract theory, microeconometrics, and experimental economics to agricultural risk management, rural development, food security, and environmental and natural resource issues in developing-country settings.

Anna Josephson is  anAssociate Professor of Agricultural and Resource Economics at the University of Arizona and Co-Director of the AIDE Lab. Her research examines measurement, data, methods, and assumptions in applied development economics, with a focus on households and small firms navigating risky environments across Sub-Saharan Africa and Asia.

Elinor Benami is an Assistant Professor of Agricultural and Applied Economics at Virginia Tech. Her research spans environmental and development economics, agriculture, and policy, with particular expertise in satellite data, machine learning, remote sensing, and the design of programs that improve environmental compliance and weather-risk management.

Patrick Behrer is an Economist in the World Bank’s Development Research Group. His research focuses on environmental and development economics, including air pollution, climate change, climate adaptation, human capital formation, and the use of big data and satellite remote sensing to study environmental conditions.

Michael J. Cecil is a postdoctoral scholar in the University of Maryland’s Department of Geographical Sciences and NASA Harvest. His research uses ground and satellite sensors, crop modeling, land surface phenology, and machine and deep learning to study crop growth, agricultural management practices, and smallholder agriculture.

Sydney Gourlay is a Senior Economist in the World Bank’s Development Data Group, working with the Living Standards Measurement Study and co-leading the Methods & Tools Development Component of the 50x2030 Initiative. Her research focuses on agricultural survey methods and methodological validation, including land area measurement, land tenure, soil health, and agricultural data quality.

Robert Heilmayr is Associate Professor of environmental and ecological economics at UC Santa Barbara’s Bren School of Environmental Science & Management, with a joint appointment in Environmental Studies. His research combines economics, geography, ecology, causal inference, and Earth observation to study conservation, land systems, deforestation, and natural resource governance.

Ella Kirchner is a Research Associate at Virginia Tech whose work focuses on agricultural index insurance and climate-risk management for smallholder farmers. Her research examines insurance adoption, farmer preferences, index design, and the use of digital technologies and Earth observation to improve resilience to drought, floods, pests, and other production risks.

Gina Maskell is a postdoctoral researcher at the Potsdam Institute for Climate Impact Research in the Adaptation in Agricultural Systems working group. Her interdisciplinary research uses geospatial and mixed quantitative social science methods to study resilience, climate adaptation, agroforestry, land-use systems, and the socioeconomic and biophysical contexts shaping agricultural responses to climate change.

Robert Heilmayr is Associate Professor of environmental and ecological economics at UC Santa Barbara's Environmental Studies Program, with a joint appointment in the Bren School of Environmental Science & Management. His research applies causal inference techniques from economics to remotely sensed data to answer policy-relevant questions about conservation and natural resource governance.

# Further Resources

| Source Type | Types of Information | Geospatial Data | Temporal Coverage | Spatial Coverage | Notes |
|---|---|---|---|---|---|
| Demographic and health Surveys | Population, health, and nutrition | (Jittered) GPS coordinates of survey clusters (not individual households) | Varies by location | Over 40 countries | Funded by U.S.AID, on pause starting January 2025. Access restricted/requires registration |
| World Bank LSMS | Mult-topic surveys on poverty and living conditions | Selected household coordinates | Varies by location | 43 countries | Access generally restricted to geospatial data.<br><br>Data and documentation available on the World Bank Microdata Library: https://microdata.worldbank.org |
| World Bank Living Standard Measurement Study - Integrated Surveys on Agriculture (LSMS-ISA) | Longitudinal, multi-topic questionnaire (household and individual living conditions, welfare, and livelihoods) with extensive data on agriculture | Selected household coordinates and plot coordinates and/or boundaries | 2008+ | Eight countries in sub-Saharan Africa (Burkina Faso, Ethiopia, Malawi, Mali, Niger, Nigeria, Tanzania, Uganda) | Anonymized cluster-level coordinates disseminated, as well as a set of geovariables for each household.<br><br>Raw geolocation data for households and plots restricted.<br><br>Data and documentation available on the World Bank Microdata Library: https://microdata.worldbank.org<br><br>https://www.worldbank.org/en/programs/lsms |
| FAO World Programme for the Census of Agriculture | | | | | https://www.fao.org/world-census-agriculture/en/<br><br>[like LSMS in some ways - repository for national surveys] |
| MICS | Children and Women's Welfare | Enumeration area coordinates | Mid 1990s + | 190+ countries | https://mics.unicef.org/surveys |
| World Bank Enterprise Surveys? | access to finance, corruption, infrastructure, crime, competition, labor, obstacles to growth, and performance measures | Business locations | 2002+ | | Firm level data access restricted.<br><br>https://www.enterprisesurveys.org/en/enterprisesurveys |
| Labor Force Surveys with Geographic Identifiers | Statistics on the labour force, employment and unemployment for monitoring | Administrative boundaries (districts, municipalities) statistics | | | Often highly aggregated, requires use of sampling weights and suggested evaluation of sampling error<br>https://ilostat.ilo.org/resources/lfs-toolkit/ |